\documentclass[conference]{IEEEtran}
\usepackage{filecontents,lipsum}
\usepackage[noadjust]{cite}
\usepackage{multicol}
\usepackage{graphicx}
\usepackage{epstopdf}
\usepackage{amsmath,amssymb}

\DeclareMathOperator*{\argmin}{arg\,min}

\begin{document}

%
\title{Performance Analysis of Split Preamble RAN Over-load Protocol for M2M Communications in Cellular Networks}
\author{\IEEEauthorblockN{Ameneh Pourmoghadas, \emph{Member, IEEE,} P. G. Poonacha}
\IEEEauthorblockA{International Institute of Information \& Technology Bangalore, India\\
Email: ameneh.langroudi@iiitb.org, poonacha.pg@iiitb.ac.in}

}

\maketitle

\begin{abstract}
Machine type communications (MTC) in 3G/4G networks is getting more attention recently due to bursty nature of traffic characteristics in contrast to Poisson type H2H traffic. A large number of methods have been suggested in the literature. In this paper we give a mathematical model on performance analysis of Disjoint Allocation (DA) and Joint Allocation (JA) methods for allocating preambles to M2M and H2H users. In an earlier work we had investigated the performance of two possible splitting preamble methods on collision probability and energy reduction for MTC subscribers. In this paper we develop a stochastic model for JA/DA method in RACH procedure using a $K$th order Markov chain approach and carry out performance analysis in terms of  collision probability, access success probability, average access delay, statistics of preamble transmissions and statistics of access delay for M2M users. The optimal number of reserved preamble set is derived based on the given success access delay threshold. Numerical results verifies the simulation results and demonstrates that this model can estimate the performance of the JA/DA method accurately.
\end{abstract}

\IEEEpeerreviewmaketitle

\section{Introduction}
Long term evolution (LTE)/LTE-Advanced and UMTS are
the radio access network (RAN) protocols for 4G
and 3G standards respectively. The study of RAN
improvements for Machine Type-Communications is still under
investigation by the 3rd Generation Partnership Project (3GPP).
This is due to the fact that machine service requirements are quite different compared to H2H services. For instance, in smart-grid networks, Advanced Metering Infrastructure (AMI) requires low
throughput and can tolerate high latency. Whereas, Automated
Demand Response (ADR) requires medium throughput and
medium to low latency and Feeder Automation (FA) requires
high throughput and low latency. Providing the promised QoS
for MTC subscribers under a massive deployment of the
number of machine devices in cellular areas is what is known
in 3GPP as RAN-overload control method. By giving
priority to the 4G cellular networks, the service providers
across the globe are looking for solutions which promises to support a wide range
of applications. Towards that end, it is important to propose and study protocols by simulation as well as mathematical analysis. A variety of RAN-overload control methods is suggested by 3GPP \cite{00} such as, extended access barring (EAB), separate PRACH (physical random access channel) resources for MTC, pull based schemes, MTC
group paging and MTC specific backoff scheme. No related
research shows the effect of the separate allocating RACH
resources on the LTE network. In this paper, the effect of
preamble splitting between H2H and M2M users is studied with 
emphasis on modeling and performance analysis. 
\\
Congestion or overload in RAN is in associated with a huge number of MTC devices attempting to establish a connection to the network radio component (eNB) specially in PRACH \footnote{This is known as LTE uplink connection process, in which devices in RRC-IDLE status contend to win a data channel from eNB and turn to RRC-CONNECTED status \cite{00-1}.}. To manage overload scenarios under the bursty arrival of the machine devices, 3GPP suggested splitting of the preamble set between H2H and M2M subscribers \cite{00}. We call it the Disjoint Allocation (DA) method. The splitting parameter will decide the performance of H2H and M2M users. However, the M2M subscribers will face more
degradation in performance as a smaller number of preambles are allocated to them.
Since the frequency of the H2H call arrival is generally much less compared to M2M arrival 
rates with bursty nature of arrivals we proposed and studied an enhanced version of the DA
method where the H2H subscribers could share their allocated preambles with the MTC devices \cite{0}. We call this approach, Joint Allocation (JA) method. In \cite{0}, we have shown through analysis and simulations that 
the JA method exhibits better performance in terms of the collision reduction and results in good energy saving for MTC devices. However, we mainly focused on the energy reduction for machine devices
and not on their latency constrains. In this paper we have given a mathematical analysis of the RACH process under JA/DA method and an optimum preamble allocation to H2H subscribers subject to providing the required access delay for M2M applications.\\
In order to get a better understanding of simulations and approximations results
of the RAN-overload control algorithms, recently there have
been interesting analytical models in literature to investigate a
couple of them. We can categorize the RAN-overload
control problems into two groups: a) resource allocation
on PRACH and PUSCH as they share fixed number of RBs,
where wasting any of the RBs imposes unused data channel
on PUSCH b) only data channel allocation on PUSCH. In this paper our focus is on case (a), in which, we assume that there are
proposed MAC procedures like \cite{00-1-1} that allows the eNB to
be informed about the collided preambles in very first steps
to prevent it allocating PUSCH RBs to them.
Papers \cite{1}, \cite{1-1} use an iterative model
to evaluate the performance of the group paging algorithm
in 3GPP with bursty arrival of MTC devices. Using mathematical models \cite{1}, \cite{1-1} find the optimum group sizes and allocated radio resources for the desired QoS
requirements of MTC devices. In \cite{2} author proves that steady state of the RACH process depends on the number of
the permitted re-transmission times under different loads of the
RACH arrival. \cite{3} uses a Markov chain model to effectively
analyze the interactions in RACH channel. Analysis of the
throughput under Poisson load in RACH procedure to examine
the impact of different RA configurations can be found
in \cite{4}. However, mathematical analysis, when the arrival of H2H and M2M
calls following Type1 and Type2 distributions as defined in
3GPP \cite{00}, or an analytical model to investigate
the performance of PRACH resource allocation
algorithms appears to be not available. 
In this letter, we provide a stochastic model and analysis for
RACH procedure based on the RAN-overload methods DA \& JA introduced in \cite{0}, which will be explained in section \ref{section IV}.\\
The remainder of this paper is
organized as follows. In section \ref{Related Works} we explain a preparation on the random access procedure in the LTE system and the related works. Section \ref{system model} elaborates the system model and the proposed stochastic model. 
Section \ref{section IV} gives a brief introduction on the proposed JA and DA algorithms and analyzes the model to evaluate the performance metrics parameters, and we derive conditions for the optimal allocated number of preambles for providing the required access success time for machine type applications. In Section \ref{section VI}, we present numerical results and the paper is concluded in Section \ref{section VII}. 
 
\section{Random Access Procedure and Related Works}
\label{Related Works}
\subsection{Random Access Procedure in LTE System}
In this subsection we provide a brief description on RACH procedure (RA), which proceed on PRACH \cite{00-1}. The RA architecture of many wireless standards including IEEE 802.11*, IEEE 802.16 (WIMAX), UMTS and LTE is similar to a
slotted ALOHA system. The only main difference between LTE and other convectional standards, is that here we are dealing with a multi-channel slotted ALOHA. Each RA-TS (time slot) in PRACH consists of multiple preambles, which are  a set of non-overlapping slots (see Figure \ref {RA}). The RA procedure in LTE is completed by the following four steps. 
\begin{itemize}
\item {Preamble transmission (step1): A UE (user equipment/subscriber) randomly selects a preamble out of all available preambles with equal probability and transmits this to the eNB.}
\item{Random access response (step2: RAR-Msg): Upon detecting preambles, the eNB sends the RAR-Msg to the UE, which is the crucial step for granting an uplink resource in next step.}
\item{Connection setup request message (step 3): When a UE
receives its RAR-Msg, by using the initially granted uplink resource from step2 the UE can send the connection setup request message.}
\item{Connection setup response message (step 4): If the eNB successfully receives the connection setup request message in step 3, the eNB sends the acknowledgement message to the UE. Upon receiving this ACK message The RA procedure is completed.}
\end{itemize}
In Figure \ref{RA}, we present an example of RA procedure described above. In this example, eight UEs arrive at $(i-1)_{th}$ RA-TS to send preamble to the eNB individually. Among them those preambles chosen by multiple UEs in this RA-TS (preamble \#2 \& \#4) are considered to be collided, and only chosen preamble \#1 and \#7 is assumed to be success\footnote{We note that there are physical layer factors such as different levels of transmission power among the UEs that can influence the probability of the success/collision of RA request. However, we focus on the behaviour of MAC layer, and thus the properties of the PHY layer is beyond the scope of this study.}. Those who send preamble without collision receive the RAR-Msg (step2) and the backlogged UEs (collided) should reattempt the procedure (we will describe the strategy of the reattempting RA procedure with details in section \ref{system model}).

\subsection{Related Works}
Since the status of the RA-TSs in RA procedure can randomly change according to the independent arrival of contending subscribers, we analyze the stochastic behaviour of the RA procedure by design a Markov model. According to the literature survey we can classify the Markov analytic works base on the state space as follows.

\begin{itemize}
\item States can represent the number of backlogged (collided) users $states=\{0,1,2,...N_{UE}\}$, in which $N_{UE}$ is the number of all users playing a contention role in RA process. The states can step forward or backward as the number of backlogged UEs can face collision or success respectfully \cite{5}, \cite{6}. 
\item States represent the number of preamble transmission $states=\{0,1,..W\}$. In which $W$ is the maximum number of allowed preamble transmission a UE can try before it gets connected to the eNB. Otherwise, UE is block to participate in the contention process\cite{1}. 
\item States in this approach represent the number of possible RA-TSs during the experiment a UE can encounter $states=\{1,2,..i,..\eta,S,d\}$. In which $\eta$ is the maximum RA-TS in the experiment, $S$ is the success state that stores all successful UEs received their success RAR Msg., and state $d$ includes the dropped UEs who exceeded their $W$ number of preamble transmission and are withdrawn from the process \cite{7}. Here, in the referenced article, author has either simplified the states' population model by omitting the impact of new arrivals and number of retransmission, or the probability of transmission is not being considered as a function of channel resources.  
\end{itemize}
In this paper we introduce a K$^{th}$ order Markov chain (K-Mc) an enhanced version of model in \cite{7} to estimate the average access success delay for M2M with separate preamble RAN-overload control method. In next section we will discuss our proposed K-Mc model in details. 

\section{System Model}
\label{system model}
\begin{figure*}[ht]
\centering
\includegraphics[width=15cm,height=4.7cm]{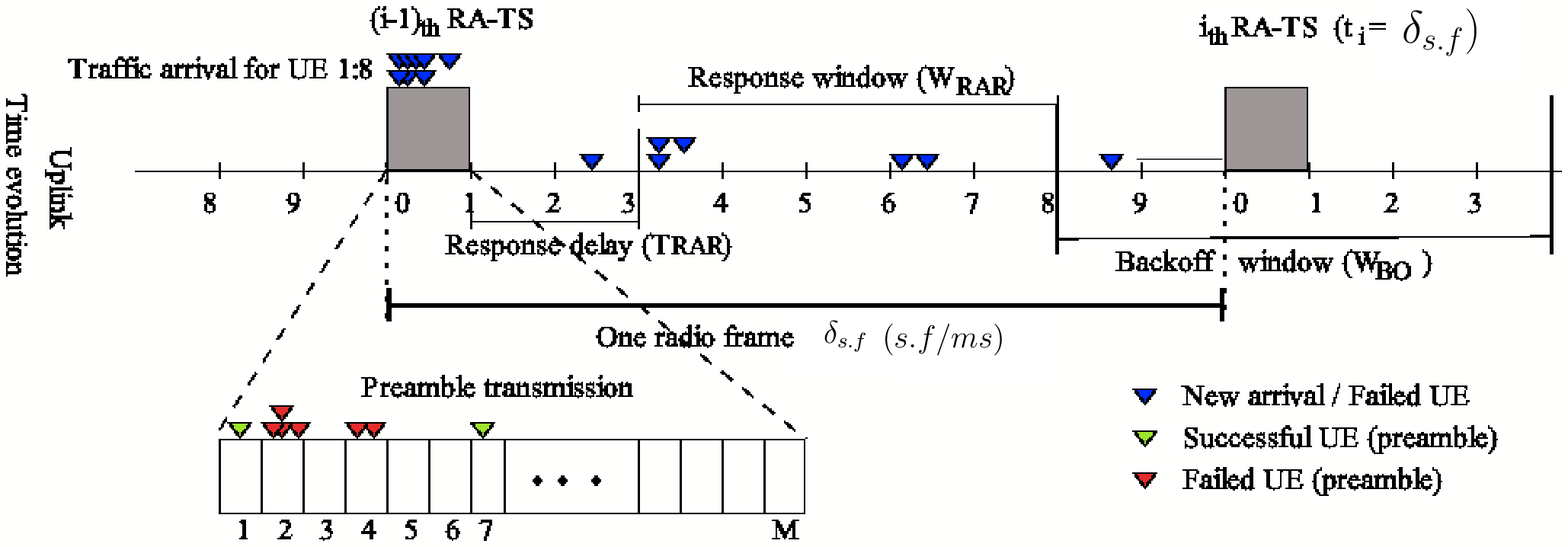}
\caption{RA procedure.}
\label{RA}
\end{figure*}
We consider a single eNB at the center of a cellular area in LTE network, in which $N_{h2h}$ and $N_{MTC}$ number of H2H and M2M subscribers are contending together. Before we start the analysis model, it is worth to discuss which mathematical model can fit on the MTC calls?  \\
In typical M2M calls, the uplink traffic dominates the downlink in comparison to the H2H calls (for instance sending regular/irregular reports). Machines pars more number of messages with short length which eventually this trend generates a bulk amount of attempts in a cellular areas. That is the reason the standard Poisson model does not meet all the characteristics of the M2M calls. In this letter, following 3GPP \cite{00} we use the traffic type2, a Beta pdf for MTCs. 
Figure \ref{RA} illustrates the time evolution of the RA procedure in uplink PRACH. Time is divided in 1ms slots representing the full duration of a sub-frame (s.f) in LTE-A. RA-TS are labeled as $ 1\leq i \leq \eta$ . We assume that the total number of $N_{h2h}+N_{MTC}$ UEs activate during $\eta$ time. When a UE fails to transmit its $n_{th} (1\leq n\leq W)$ RA request (unsuccessful preamble) at time slot $(i-1)_{th}$ it repeats the RA procedure and reattempts in future upcoming RA-TS ($i, i+1, i+2 ...$). Here $n=1$ refers a new arrival UE who is attempting the RA procedure as first time and $n=W$ refers to a failed UE for $W-1$ times, which is reattempting the RA procedure for the last permitted time. Therefore arrival in each RA-TS can be a composition of new and backlogged RA arrival of H2H and M2M calls.\\ If there is a new arrival ($n=1$) UE departing between two RA-TS it waits and attempts its preamble transmission on immediate next RA-TS. For instance, in this example seven new arrival UEs who departure between $(i-1)_{th}$ and $i_{th}$ RA-TSs will attempt their first preamble transmission on $i_{th}$ RA-TS. After sending the preamble message UE should wait upto maximum $T_{RAR}+W_{RAR}$ time (RAR delay response and RAR window) to receive the RAR-Msg. In this example two UEs who contending in $(i-1)_{th}$ RA-TS and picked the preamble \#7 \& \#1 will receive their RAR-Msg after 7ms. Upon receiving the RAR-Msg UE is considered to be success and its total delay is the variance $(t_{j_{th}}+T_{RAR}+W_{RAR})-t_{k}$. In which, $t_{k},\ t_{j_{th}}$ resemble the time of UE's initial arrival in $k_{th}$ TS (note that it can be a RA-TS or another uplink TS), and the successful $j_{th}$ RA-TS attempt. The rest of UEs who picked up a same preamble won't be able to receive their RAR-Msg are failed. Failed UEs after $T_{RAR}+W_{RAR}$ time will start back-off procedure, by choosing uniformly a random number between [0-$W_{BO}$]. UE starts counting down, upon its timer is expired, it starts reattempting $n $th RA procedure $(2\leq n \leq W)$ in upcoming RA-TS. Details of all notations used in this paper is given in Table \ref{t_parameter}. This process continues till $t_{\eta}$. We are interested in estimating the average access success delay for UEs which is the interval between the first RA attempt and the RA procedure completion that will be explained in following sections\footnote{One should note that in this paper the delay of PHY layer such as processing time from eNB or UE end and contention resolution time-window is omitted. Also as discussed earlier we have assumed that eNB can recognize about the fail/success status of preamble requests at Msg2, therefore the the delay of receiving Msg3-Msg4 is not considered in  our computation.}.
A K-Mc model is introduced in following subsection.  
\subsection{Framework/Problem formulation}
\begin{figure}[t]
\centering
\includegraphics[scale=0.30]{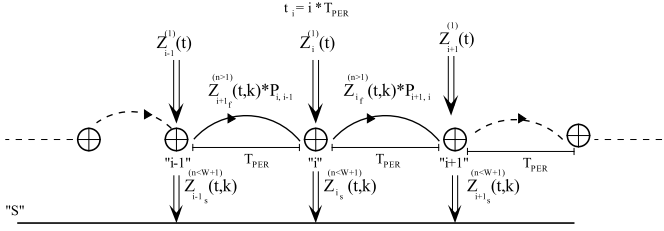}
\caption{A K-Mc demonstration for RACH stochastic process.}
\label{KMC}
\end{figure}

We are interested in finding a probability for collision and success transmission for each RA-TS that leads us in computing the expected access success delay and other performance metrics through analysis. By considering this fact that the number of contending users (M2M or H2H) per RA-TS is stochastic by nature, we denote them with the random variable, $Z^{(n)}_{i}(t,k_{i})$ as depicted in Figure \ref{KMC}. In which, $Z^{(n)}_{i}(t,k_{i}), 2\leq n \leq W$ is the expected number of arrival at given time $t$ in $i^{th}$ RA-TS after $(n-1)^{th}$ preamble re-transmission in past $t-k_{i}$ time, and its transactions can be modeled by a K$^{th}$ order Markov chain which we are going to discuss in this section. \\
As it is deduced from the Figure \ref{KMC} $Z^{(n)}_{i}(t,k_{i})$ is determined by the composition of the number of new arrivals $Z^{(1)}_{i}(t)\ (n=1)$ and number of failed users $Z^{(n)}_{i_{f}}(t,k_{i}),\ 2\leq n \leq W $ who come from the $i-1^{th}, i-2^{th} ... i-k^{th}$ previous RA-TSs. The cross symbols in this diagram represent the state space\footnote{Sate space is resembling the possible RA-TSs in RA procedure, therefore we use them interchangeably.} $S=\{ Z_{1}(t,k_{1}), Z_{2}(t,k_{2})... Z_{i}(t,k_{i}),...Z_{\eta}(t,k_{\eta}), Z_{S}(t)\}$, where $1 \leq i \leq \eta$, $Z_{S}(t)$ is the success state that consist of successful UEs at time $t$ and the sequences of time $t_{1}\leq t_{2}\leq t_{3}...\leq T$ are considered to be 1ms and $T$ is the latest duration of the users activation during the experiment. $Z_{i}(t,k_{i})$ is the expected number of users arrival at time $t$ in $i^{th}$ RA-TS in total, which is defined as
\begin{equation}\label{zz_{i}}
Z_{i}(t,k_{i})=\sum_{n=1}^{W}Z_{i}^{(n)}(t,k_{i}).
\end{equation}
To clarify the $k$ step transmission in our model let us recap the back-off technique in RA procedure. Let's assume at time $t$, $Z_{1}^{(1)}(t)$ number of new arrivals departure in state $1$. without contributing the failed UEs from previous states, these population of users will contend together on $M=\{m_{1},m_{2},...M \}$ number of preambles which are available in this state. The second order failed UEs who are failed with their first preamble transmission will re-attend their luck in upcoming states $(1+1, 1+2...)$, and if their transmission failed again, they will reattempt their third luck and so on. UEs may continue this process of re-transmission up to $W$ times. The conditional probability that the stochastic RACH state (RA-TS) $j$ at time $t-m$ will switch to state $i$ at time $t$ is
\begin{align}
P_{i,j} &=\{ X(t)=\mathop{\mathbb{E}}[Z_{i}(t,k_{i})]|X(t-m)=\mathop{\mathbb{E}}[Z_{i-m}(t,k_{i-m})]\nonumber ,\\\nonumber 
&m=1,2,..k_{i};\ 2\leq i\leq \eta\}>0,\\ 
&\sum_{j=i-1}^{k_{i}} P_{i,j}=1.
\end{align}
The population of arrival in each future states depends on $\delta_{s.f}$ the period between two sequential RA-TS and the size of the back-off window $W_{BO}$. For instance, the maximum number of past possible steps from $i^{th}$ state that can associate in backlogged arrival process in this state can be defined as
\begin{equation}\label{k}
k_{i}=max\{\Delta t_{i,j}\setminus \delta_{s.f},\ j=1,2,...i-1 \},
\end{equation}
in which $t_{i}=i*\delta_{s.f}$ ($i\in \{1,2,...\eta\}$), and $\Delta t_{i,j}= t_{i}-t_{j}$. After $T_{RAR}+W_{RAR}$ time, a failed UE in any state will chose a number between $[1-W_{BO}]$ with equal probability $\frac{1}{W_{BO}}$. So we can represent the probability of transmission from state  $j$ to the state $i$ as following
 \begin{equation}\label{pi,j}
  P_{i,j}=
    \begin{cases}
      \frac{\delta_{s.f}}{\Delta W_{BO}}, & \text{if}\ a>1\ \& \ Bo\_max_{j}\geq t_{i} \\
      \frac{Bo\_max_{j}-(t_{i}+1)}{\Delta W_{BO}}, & \text{if}\ a>1\ \& \ t_{i-1}\leq Bo\_max_{j}\leq t_{i}\\
      \frac{(t_{i}+1)-Bo\_min_{j}}{\Delta W_{BO}}, & \text{if}\ a\leq 1\ \& \ Bo\_max_{j}\geq t_{i}\\
      0, & \text{otherwise}
    \end{cases}.
  \end{equation}
In which the parameters related to each state is defined as \\
$a=\Delta t_{i,j}\setminus \delta_{s.f},\ j=1,2,...i-1$,\\
$Bo\_min_{j}=t_{j}+T_{RAR}+W_{RAR}$,\\
$Bo\_max_{j}=Bo\_min_{j}+W_{BO}$,\\
$\Delta W_{BO}=(Bo\_max_{j}-Bo\_min_{j})+1$.

We will explain the usage of transmission probability in an example in this section. \\
As it comes from Eq. (\ref{pi,j}), loop back or backward steps in this K-Mc model is not possible. Having $P_{i,j}$ function, we can define the transition probability matrix of $(\eta*\eta)$ size as follows 
\begin{eqnarray}
[P_{i,j}] &=&  \begin{bmatrix}
0	&0	&0 &...& &0\\
P_{2,1}	&0	&0	&...& &0\\
P_{3,1}&P_{3,2}&0&0&...&0\\ 
P_{4,1}&P_{4,2}&P_{4,3}&0&...&0\\ 
				.\\
				.\\
P_{\eta,1}&P_{\eta,2}&P_{\eta,3}&&...&P_{\eta,\eta-1}				               
        \end{bmatrix}.
\end{eqnarray}
Where $2\leq i\leq \eta$ and $1\leq j\leq i-1$. Having the $P_{i,j}$ matrix we can estimate the expected number of arrivals at time $t$ in $i^{th}$ state as defined in Eq. (\ref{zz_{i}}) where,
\begin{equation}\label{z_{i}}
Z_{i}^{(n)}(t,k_{i})=\sum_{j=i-1}^{k_{i}}P_{i,j}*Z_{j_{f}}^{(n-1)}(t,k_{j}),\ 2\leq n\leq W,
\end{equation}
\begin{equation}\label{Z}
\Rightarrow Z_{i}(t,k_{i})=Z_{i}^{(1)}(t)+\sum_{n=2}^{W}\sum_{j=i-1}^{k_{i}}P_{i,j}*Z_{j_{f}}^{(n-1)}(t,k_{j}).
\end{equation}
In which $Z_{j_{f}}^{(n)}(t,k_{j})$ is the expected number of failed UEs arrival from state $j$ who are attempting for $n^{th}$ re-transmission in current state $i$, and their population is defined as
\begin{equation}
Z_{j_{f}}^{(n)}(t,k_{j})=Z_{j}^{(n)}(t,k_{j})*P_{j_{f}},\ 2\leq n\leq W.
\end{equation} 
$P_{j_{f}}$ is the failing probability in state $j$. Let us define success probability $P_{i_{s}}$ as the probability that exactly each preamble $m_{i}$ in $i_{th}(1\leq i \leq \eta)$ state is chosen by one user, given the expected number of arrivals, $Z_{i}$, as
\begin{equation}\label{P_s}
P_{i_{s}}=P\{m_{i}=1|X_{i}(t,k)=Z_{i},\  m_{i} \in \{1,2...M \} \},
\end{equation}
so we can write
\begin{equation}\label{P_f}
P_{i_{f}}=P\{m_{i}>1|X_{i}(t,k)=Z_{i},\  m_{i} \in \{ 1,2...M \}\}=1-P_{i_{s}}.
\end{equation}
Perhaps expanding Eq. (\ref{zz_{i}}) worth to practice before we start deriving the other parameters. As it comes from Eq. (\ref{z_{i}}) the $n^{th}$ arrival at time $t$ in current state is depend on failed population UEs from $(n-1)^{th}$ time preamble attempt in previous states. For example, if the system at time $t$ is in state $i=4$, given $\delta_{s.f}=10ms,\ \Delta W_{BO}=21ms$, from Eq. (\ref{k}) we can compute the number of associated past states $k_{4}=3$ and their transition probability $P_{4,j},\ (j=1,...3)$ as: $P_{4,1}=0.1,\  P_{4,2}=0.5,\  P_{4,3}=0.4$. Using Eq. (\ref{z_{i}}),\ (\ref{Z}) the expected number of arrivals in this state which is in associated with the new arrival $Z_{4}^{(1)}(t)$ and failed users in previous states 1,\ 2 \& 3 can be denoted as
\begin{equation}
Z_{4}(t,k_{4}) =Z_{4}^{(1)}(t)+Z_{4}^{(2)}(t,k_{4})+Z_{4}^{(3)}(t,k_{4})+Z_{4}^{(4)}(t,k_{4}),
\end{equation}
in which,
\begin{align}
Z_{4}^{(1)}(t) &=Z_{4},\\ \nonumber
Z_{4}^{(2)}(t,k_{4}) &=P_{4,1}*Z_{1_{f}}^{(1)}(t,k_{1})+P_{4,2}*Z_{2_{f}}^{(1)}(t,k_{2})\\ \nonumber
 & +P_{4,3}*Z_{3_{f}}^{(1)}(t,k_{3}),\\ \nonumber
Z_{4}^{(3)}(t,k_{4}) &=P_{4,2}*Z_{2_{f}}^{(2)}(t,k_{2})+P_{4,3}*Z_{3_{f}}^{(2)}(t,k_{3}),\\ \nonumber
Z_{4}^{(4)}(t,k_{4}) &=P_{4,3}*Z_{3_{f}}^{(3)}(t,k_{3}).
\end{align}
Where, $Z_{4}$ is the contribution of initially distributed users in $4^{th}$ state at time $t$. Note that the rest of higher order failed UEs ($n>4$) are not possible in this state. As it comes from the above derivation we can easily compute the other states population recursively.
\begin {table}[t]
\caption{Parameters}
\begin{tabular}{|c|c|}
\hline
\textbf{Parameter} & \textbf{Value}  \\ \hline
M (total number of preambles/RA-TS)	   & 54  \\ \hline
$T_{RAR}$(Msg2 response delay) & 2 ms\\ \hline
$W_{RAR}$ (Msg2 response window) & 5 ms\\ \hline 
$\delta_{s.f}$ (RACH sub frame interval)& \begin{tabular}[x]{@{}c@{}}10 ms\\if PRACH Configuration Index = 6
\end{tabular} \\ \hline
$W$ (maximum transmissions) & 10 \\ \hline
$W_{BO}$ (Backoff window) & 20 ms\\ \hline
$T$ (latest duration of the users activation)  & 10sec \\ \hline
$\eta$ (last Markov state / RA-TS )& $\frac{T}{\delta_{s.f}}$ \\ \hline
H2H and M2M call probability distribution function& Type1 \& Type2 \cite{00}\\ \hline
\end{tabular}
    
    \label{t_parameter}
\end{table}
\subsection{Performance Metrics Analysis in RACH Procedure}
In this subsection we analyze the performance metrics in RACH process which is in associated with two types of users H2H and M2M. To begin with, we start with clear definition of performance metrics specified in 3GPP TR 37.868 as follows \cite{00}. 
\begin{enumerate}
\item \textbf{Access success probability}, defined as the ratio between the number of successful users who complete the RA procedure within $W$ number of preamble transmissions and the number of active users. Which should not be mistaken by the probability of success in respect to the RAOs (random access opportunity) as worked in \cite{1}, \cite{8}. It can be written as
\begin{equation}\label{asp}
P_{s}=\frac{\text{No.\ of\ success\ UEs\ within\ $W$\ retry}}{\text{Total\ no.\ of\ active\ UEs}}.
\end{equation}
In this work we have assumed the success preamble in step2 as a successful user. For the corresponding number of contending MTC, $Z_{i}^{M2M}(t,k_{i})$, and human type call arrivals, $Z_{i}^{H2H}(t,k_{i})$, at time $t$ in state $i$, the number of success UEs can be defined as
\begin{equation}
Z_{i_{s}}(t,k_{i})=P_{i_{s}}^{H2H}*Z_{i}^{H2H}(t,k_{i})+P_{i_{s}}^{M2M}*Z_{i}^{M2M}(t,k_{i}).
\end{equation}
In which, the probability of success for H2H and M2M communication in state  $i$ is defined as \cite{0}
\begin{align}\label{P_SH2H}
& P_{i_{s}}^{H2H} \{m_{i} =1|X_{i}(t,k_{i})=\mathbb{E}[Z_{i}^{H2H}(t,k_{i})], \nonumber\\
& m_{i}\in\ \{1,2...M\} \} \nonumber\\
&=\ exp(\frac{-\mathbb{E}[Z_{i}^{H2H}(t,k_{i})]}{M}),
\end{align}
\begin{align}\label{P_SM2M}
& P_{i_{s}}^{M2M} \{ m_{i} =1|X_{i}(t,k_{i})=\mathbb{E}[Z_{i}^{M2M}(t,k_{i})], \nonumber\\
& m_{i}\in\ \{1,2...M\}\}\nonumber\\
&=\ (1-\frac{1}{M})^{\mathbb{E}[Z_{i}^{M2M}(t,k_{i})]-1}.
\end{align}
Therefore, the access success probability Eq. (\ref{asp}) can be computed as 
\begin{equation}
P_{s}=\frac{\sum_{i=1}^{\eta}\sum_{n=1}^{W}Z_{i_{s}}^{{H2H}(n)}(t,k_{i})+Z_{i_{f}}^{{M2M}(n)}(t,k_{i})}{\sum_{i=1}^{\eta}Z_{i}(t,k_{i})},
\end{equation}
where,
\begin{equation}
Z_{i}(t,k_{i})=\sum_{n=1}^{W}Z_{i}^{H2H(n)}(t,k_{i})+Z_{i}^{M2M(n)}(t,k_{i}).
\end{equation}
\item \textbf{Collision probability}, defined as the ratio between the number of collided users and the total number of RAOs (with or without access attempts). If two (or more) users select the same preamble at the same RA-TS, the eNB will not be able to decode any of the preamble requests; hence, eNB will not send RAR msg. \\
Having $RAOs=\eta*M$, by using Eq. (\ref{P_f}) and Eq. (\ref{P_SH2H})-(\ref{P_SM2M}) we can derive the collision probability as
\begin{equation}
P_{f}=\frac{\sum_{i=1}^{\eta}\sum_{n=1}^{W}Z_{i_{f}}^{{M2M}(n)}(t,k_{i})+Z_{i_{f}}^{H2H(n)}(t,k_{i})}{\eta*M},
\end{equation}
in which 
\begin{align}
& Z_{i_{f}}^{{H2H}(n)}(t,k_{i})=\\\nonumber 
&Z_{i}^{{H2H}(n)}(t,k_{i})*(1-exp(\frac{-Z_{i}^{H2H}(t,k_{i})}{M})),
\end{align}
\begin{align}
& Z_{i_{f}}^{{M2M}(n)}(t,k_{i})=\\\nonumber 
& Z_{i}^{{M2M}(n)}(t,k_{i})*(1-(1-\frac{1}{M})^{Z_{i}^{M2M}(t,k_{i})-1}).
\end{align}
\item\textbf{ Statistics of number of preamble transmissions}, defined as the CDF of the number of successful users who connect to the eNB within $r\ (r\leq W)$ number of preamble retransmission to the total number of successful users. Which can be written as
\begin{equation}
F_{p}(p\leq r)=\frac{\sum_{i=1}^{\eta}\sum_{n=1}^{r} Z_{i_{s}}^{(n)}(t,k_{i})}{\sum_{i=1}^{\eta}\sum_{n=1}^{W} Z_{i_{s}}^{(n)}(t,k_{i})}.
\end{equation}
\item \textbf{Access success delay}, defined as the expected delay for successful users who complete the RA procedure within $W$ number of retransmission preamble to the total number of successful users. Which can be denoted as
\begin{equation}
\mathbb{E}[\tau]=\frac{\sum_{i=1}^{\eta}\sum_{n=1}^{W} \overline{\tau_{i}}^{(n)}}{\sum_{i=1}^{\eta}\sum_{n=1}^{W} Z_{i_{s}}^{(n)}(t,k_{i})},
\end{equation}
in which, $\overline{\tau_{i}}^{(n)}$ is the average success delay at $n^{th}$ retransmission in state $i$. Basically users delay comprises of three terms\\
$\text{User's delay}=\text{delay}_{stochastic\ RACH\ proc.}+\text{delay}_{barring}+\text{delay}_{departure}$.\\
Without considering the factor of barring delay, let's consider those UEs who failed in previous $k_{i}$ states and their $n^{th}$ preamble transmission is successful in state $i$. We can write the expected access success delay for successful users in $i^{th}$ state as
\begin{equation}
\overline{\tau_{i}}=\overline{\tau_{i}}^{(1)}+\overline{\tau_{i}}^{(2\leq n \leq W)},
\end{equation}
where the expected success access delay for new arrivals ($\overline{\tau_{i}}^{(1)}$ delay of departure), in state $i$ can be estimated as expected number of success UEs in this state multiplied by their average delay for departing between $i-1$ and $i^{th}$ state, ($\delta_{s.f}/2$), and delay of receiving RAR message, ($T_{RAR}+W_{RAR}$) as
\begin{align}
\overline{\tau_{i}}^{(1)} &= Z_{i}^{(1)}(t)*P_{i_{s}}\left \{X_{i}(t)=Z_{i}^{(1)}(t)\right\}\nonumber\\
& *(\frac{\delta_{s.f}}{2}+T_{RAR}+W_{RAR}), \label{tau1}
\end{align}
and the average access delay for higher order preamble retransmissions in this state can be estimated as 
\begin{align}
\overline{\tau_{i}}^{(2\leq n\leq W)} &= \sum_{j=i-1}^{k_{i}}Z_{j_{f}}^{(n-1)}(t,k_{j})*P_{i,j}\nonumber\\
& * P_{i_{s}}\left \{X_{i}(t)=Z_{j_{f}}^{(n-1)}(t,k_{j})*P_{i,j}\right\}\nonumber \\
& *(k_{j}*\delta_{s.f}+T_{RAR}+W_{RAR}).\label{taun}
\end{align}
In which, term $k_{j}*\delta_{s.f}$ is an extra delay for those UEs who collided in previous $j_{th}$ state ($k_{i}\leq j\leq i-1$) and re-attempt in state $i$. 
\item \textbf{Statistics of access delay}, defined as the CDF of the delay for successful users who complete the RA procedure within $\omega$ number of retransmission time divided by total number of successful users. Which can be denoted as
\begin{equation}
F_{\tau}(\tau\leq \omega)=\frac{\sum_{i=1}^{\eta}\sum_{n=1}^{\omega} \overline{\tau_{i}}^{(n)}}{\sum_{i=1}^{\eta}\sum_{n=1}^{W} Z_{i_{s}}^{(n)}(t,k_{i})}.
\end{equation}
\end{enumerate}
\section{Splitting RACH Resources For M2M And H2H Communication And Parameter Optimization For Proposed JA Scheme}\label{section IV}
According to \cite{00} when MTC and H2H calls share the RACH resources, they experience the same access collision probability. Providing separate RACH resources for the H2H and MTC devices (DA method) can reduce the number of collisions, however as we discussed in \cite{0} JA method has better performance in reducing the collision rate for MTC calls in comparing to the DA algorithm. In this section we briefly describe the mentioned preamble splitting RAN-overload control algorithms, and then we develop the access success delay performance metrics through analysis. The target in this section is to find an optimum number of preamble allocation to H2H calls, $x_{i}^{\dagger},\ (1\leq i\leq\eta)$, in order to maintain the required access success delay for machine type applications below a given threshold using JA method. 
\begin{figure}[tp]
\centering
\includegraphics[scale=0.35]{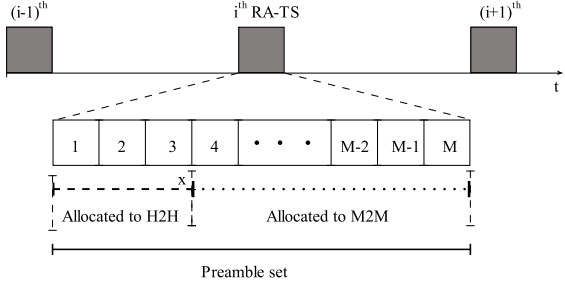}
\caption{Scheme of JA \& DA splitting preamble methods in RACH procedure.}
\label{JA_DA}
\end{figure}
\subsection{DA and JA RAN-Overload Control Methods}
As it's depicted in Figure \ref{JA_DA} each state $i$ contains $M$ preambles $M=\{m_{1},m_{2},...M\}$ shared between human and machine type devices. In DA algorithm, users have access to a fixed number of allocated preambles and contend. For example, in this figure $x=a$ and $M-a$ is the number of preambles allocated to the H2H and M2M communication respectively. Whereas, in JA algorithm, H2H users have access to $x$ number of preambles ($x$ is a random variable) and M2M users contend with them using the access permission to the whole $M$ preamble set. Which means that, the $x$ number of preambles will be shared between M2M and H2H, whereas still these are M2M users who have access to the $M-x$ of them. Let $\lambda_{i}$ be the average number of H2H new arrivals in state $i$, $g(t)$ be the probability distribution function of RACH requests generated by the M2M calls, and $N_{MTC}$ be the total number of M2M active users during $[0-T]$ time. The number of M2M new arrivals sending their preamble request in state $i$ can be defined as defined in \cite{00}.
\begin{equation}
N_{i}=\ N_{MTC}\int_{t_{i}}g(t)\ dt, 
\end{equation}
We can write the expected number of new arrivals in $i^{th}$ RA-TS ($i\in \{1,2,...\eta \}$) with respect to DA and JA method respectively as follows
\begin{align}\label{method1}
Z\_DA_{i}^{(1)}(t)&=\lambda_{i}|_{a}+N_{i}|_{M-a},
\end{align}
in which $\lambda_{i}$ and $N_{i}$ users contend for $a$ and $M-a$ number of preambles individually, and
\begin{align}\label{method2}
 Z\_JA_{i}^{(1)}(t)&=(\lambda_{i}+N_{i}\frac{x}{M})|_{x}+(N_{i}\frac{M-x}{M})|_{M-x},
\end{align}
where $(\lambda_{i}+N_{i}\frac{x}{M})$ number of initial users share $x$ number of preambles together. Therefore, we can re-write the population of $i^{th}$ state in general (either for DA or JA method) as follows.
\begin{align}\label{}
&Z\_DA[JA]_{i}(t,k_{i})=Z\_DA[JA]_{i}^{(1)}(t)+\\\nonumber
&\sum_{j=i-1}^{k_{i}}\sum_{n=2}^{W}(Z_{j_{f}}^{H2H(n-1)}(t,k_{j})+Z_{j_{f}}^{M2M(n-1)}(t,k_{j}))*P_{i,j}.
\end{align}
One should notice that we can find the expected number of arrivals by considering this fact that even though the new arrival of M2M users follow a Beta distribution, yet their higher orders in back-off procedure follows the uniform distribution. 
Therefore, this is easy to calculate the performance metrics with respect to the DA/JA method, that we neglect to write here. 
\subsection{Preamble Allocation Optimization for M2M Calls Given the Access Success Delay Threshold \& Number of Preambles per RA-TS}
An optimum value for $x_{i}^{\dagger}$, that minimizes the access success delay for M2M calls using JA algorithm for a given maximum threshold delay $\phi$, that can be denoted as  
\begin{equation}\label{ta_opt1}
min \{ \mathbb{E}[\tau_{JA_{i}}^{M2M}(x^{\dagger}_{i},M)]\}\ \text{with respect to $x_{i}$},
\end{equation}
where,
\begin{align}\label{ta_opt}
&\mathbb{E}[\tau_{JA_{i}}^{M2M}(x^{\dagger}_{i},M)]=\\\nonumber
&\mathbb{E}[\frac{\sum_{n}Z_{i}^{M2M(n)}(t,k_{i})*P_{i_{s}}^{M2M}*\tau_{i}}{\sum_{n}Z_{i}^{M2M(n)}(t,k_{i})*P_{i_{s}}^{M2M}+Z_{i}^{H2H(n)}(t,k_{i})*P_{i_{s}}^{H2H}}]=\\\nonumber
&\mathbb{E}[\tau_{i}]\mathbb{E}[\frac{\sum_{n}Z_{i}^{M2M(n)}(t,k_{i})*P_{i_{s}}^{M2M}}{\sum_{n}Z_{i}^{M2M(n)}(t,k_{i})*P_{i_{s}}^{M2M}+Z_{i}^{H2H(n)}(t,k_{i})*P_{i_{s}}^{H2H}}],
\end{align}
in which from Eq. (\ref{taun}), we can write 
\begin{align}
\mathbb{E}[\tau_{i}] =(\frac{k_{i}(k_{i}+1)}{2})*\delta_{s.f}+T_{RAR}+W_{RAR}
\end{align}
and,
\begin{align}\label{convex}
P_{i_{s}}^{M2M}=&(1-\frac{1}{x_{i}})^{Z_{i}^{H2H}(t,k_{i})+Z_{i}^{M2M}(t,k_{i})*(\frac{x_{i}}{M})-1}\\\nonumber
&+(1-\frac{1}{M-x_{i}})^{Z_{i}^{M2M}(t,k_{i})*(\frac{M-x_{i}}{M})-1}\\\nonumber
& P_{i_{s}}^{H2H}=exp(\frac{-Z_{i}^{H2H}(t,k_{i})}{x_{i}}).
\end{align}
Eq. (\ref{ta_opt1}) is too complex to use. Let's denote optimum $x^{\dagger}_{i}$ such that 
\[ \argmin_{x_{i}} \{ \mathbb{E}[\tau_{JA_{i}}^{M2M}(x^{\dagger}_{i},M)]\} =min\{ \mathbb{E}[\tau_{i}]A(x_{i}^{\dagger},M),B^{-1}(x_{i}^{\dagger},M) \}\]
where the two convext functions $A,\ B$ are the numerator and denominator in Eq. (\ref{ta_opt}).
Using the Jensen's inequality for any $\Phi$ convex function and random variable $X$ we have
\begin{equation}
\Phi(\mathbb{E}[X])\leq \mathbb{E}[\Phi(X)],
\end{equation}
let's assum the expected number of M2M and H2H new arrivals in state $i$ is $N_{i}$ and $\lambda_{i}$ respectively. So we can write 
\begin{align}\label{A}
&A_{i}(x_{i},M)\geq\\\nonumber
&\alpha_{i}(x_{i})(1-\frac{1}{x_{i}})^{\alpha_{i}(x_{i})-1}
+\gamma_{i}(x_{i})(1-\frac{1}{M-x_{i}})^{\gamma_{i}(x_{i})-1}, 
\end{align}
and 
\begin{align}
B_{i}(x_{i},M)&\geq A_{i}(x_{i},M)+\lambda_{i}exp(\frac{-\lambda_{i}}{x_{i}}).
\end{align}
Where $\alpha_{i}(x_{i})=\lambda_{i}+N_{i}\frac{x_{i}}{M}$, $\gamma_{i}(x_{i})=N_{i}\frac{M-x_{i}}{M}$. The R.H.S of the Eq.(\ref{A}) is the minimum expected value for function $A$. Without loosing the generality, by finding the minimum value for the function $A$ and maximum value for function $B$ we can minimize the whole fraction. To maximize the function $B$ we can let the system to reach its steady state in which the number of arrivals per RA-TS will reach to $W$ times of its new arrivals. Therefore, by numerically solving the following equation we can obtain $x^{\dagger}_{i}$ with respect to $\phi$ as 
\begin{equation}
solve\{ \frac{R.H.S[A(x^{\dagger}_{i};M,\lambda_{i},N_{i})]}{R.H.S[B(x^{\dagger}_{i};M,W*\lambda_{i},W*N_{i})}=\phi , x^{\dagger}_{i}\}.
\end{equation}
\section{Numerical Results}\label{section VI}
\begin{figure}[t]
\centering
\includegraphics[scale=0.60]{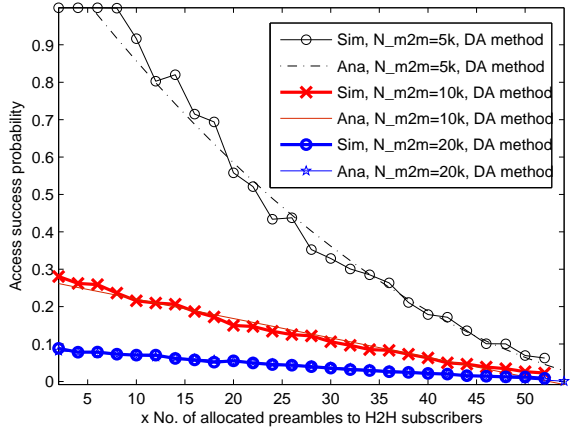}
\caption{M2M access success probability with DA method ($P_{s_{DA}}^{M2M}$), with $\lambda_{h2h}=0.5$.}
\label{PR_success_DA}
\end{figure}

\begin{figure}[t]
\centering
\includegraphics[scale=0.62]{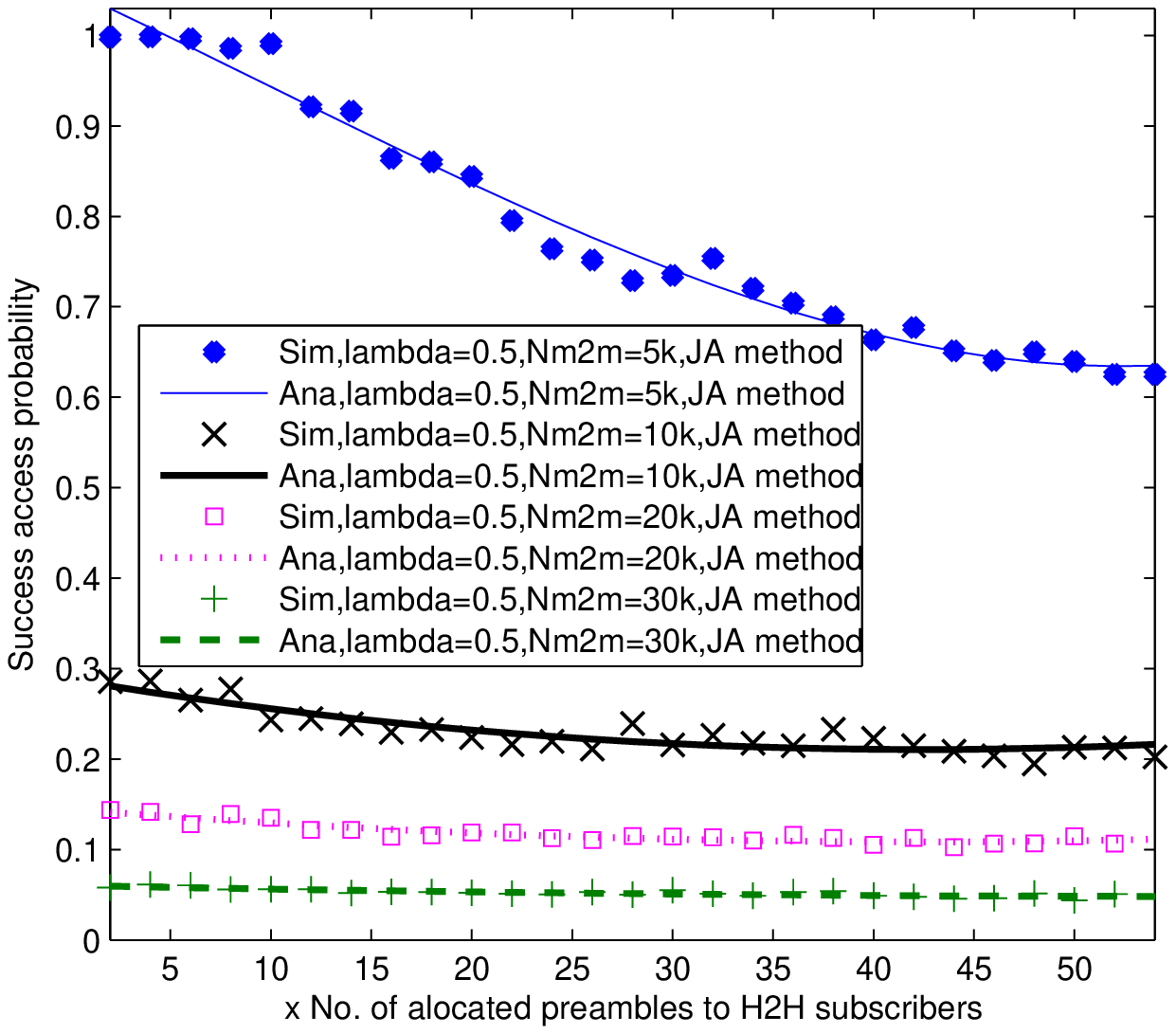}
\caption{M2M access success probability with JA method ($P_{s_{JA}}^{M2M}$), with $\lambda_{h2h}=0.5$.}
\label{PR_success_JA}
\end{figure}

\begin{figure}[t]
\centering
\includegraphics[scale=0.60]{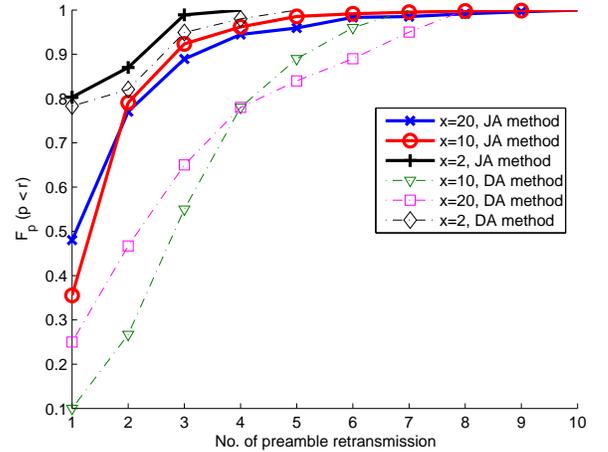}
\caption{CDF of success preamble trasnmission; $\lambda_{h2h}=0.5$.}
\label{preamble_cdf}
\end{figure}

\begin{figure}[t]
\centering
\includegraphics[scale=0.65]{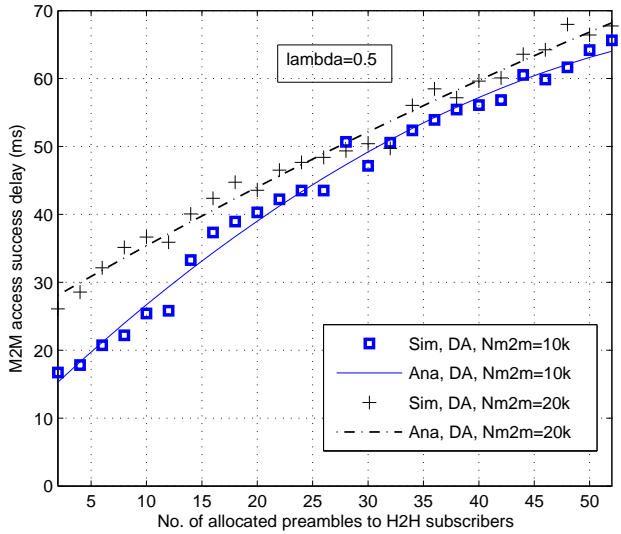}
\caption{Access success delay for M2M calls using DA method, ($\tau^{M2M}_{DA}$).}
\label{delay_DA}
\end{figure}

In this section, we aim to show the accuracy of RACH performance metrics approximation using the proposed K-Mc model in comparison to the simulation results. The analysis results include the optimal number of preamble allocation to H2H calls (i.e., $x^{\dagger}$) in PRACH, in order to avoid excess success access delay for machine type applications. Under the proposed JA method, we compare the analytical results with simulation to see for how many $N_{MTC}$ of simultaneous machine devices in a cell, the system can support the access delay requirements for M2M calls.
\begin{figure}[t]
\centering
\includegraphics[scale=0.50]{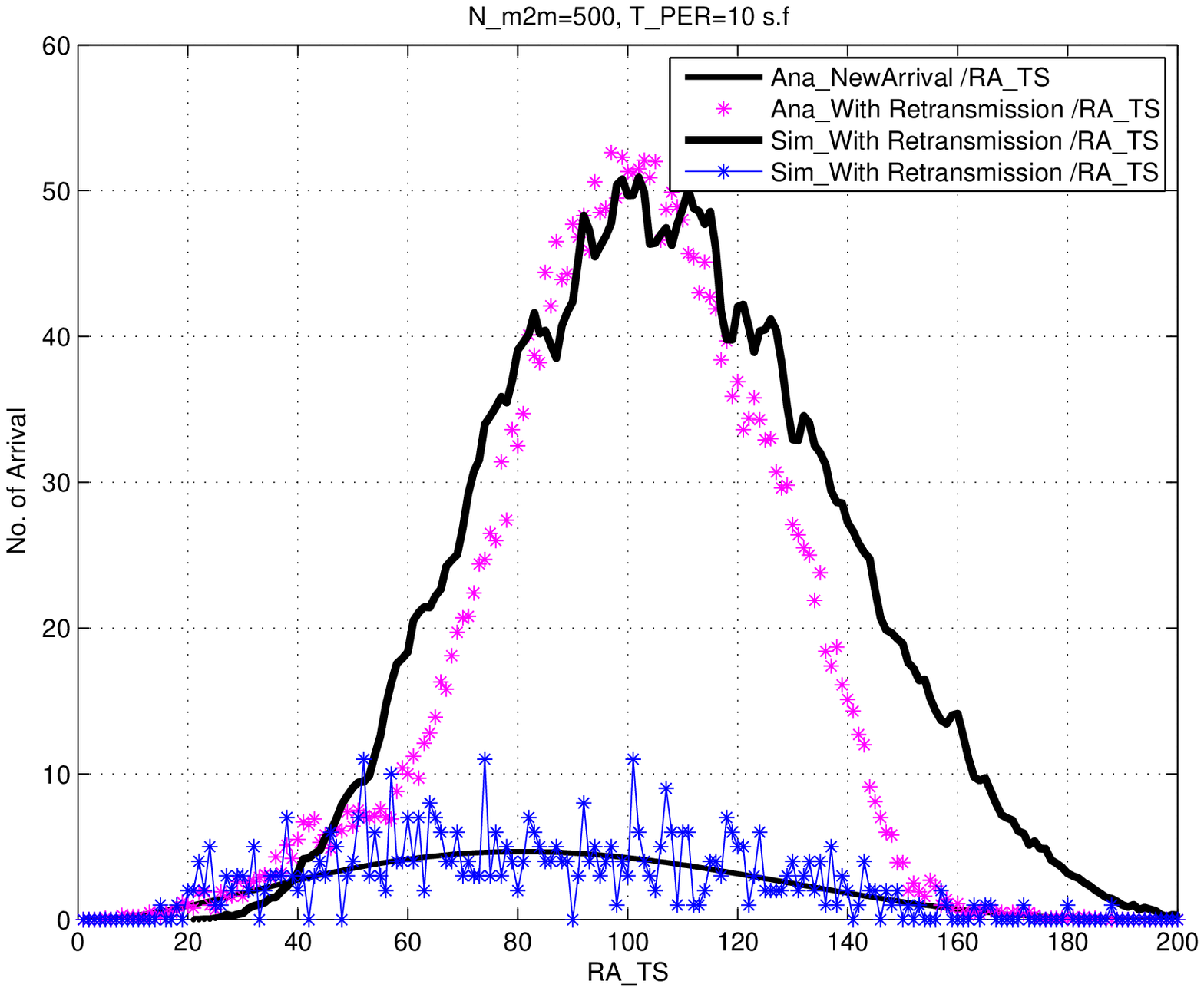}
\caption{Estimated number of M2M UEs arrival after retransmission ($Z_{i}^{M2M(n)}$) and thier new arrival ($Z_{i}^{M2M(1)}$) 
vs. simualtion ($W=10$).}
\label{Z_m2m}
\end{figure}
\begin{figure}[t]
\centering
\includegraphics[scale=0.65]{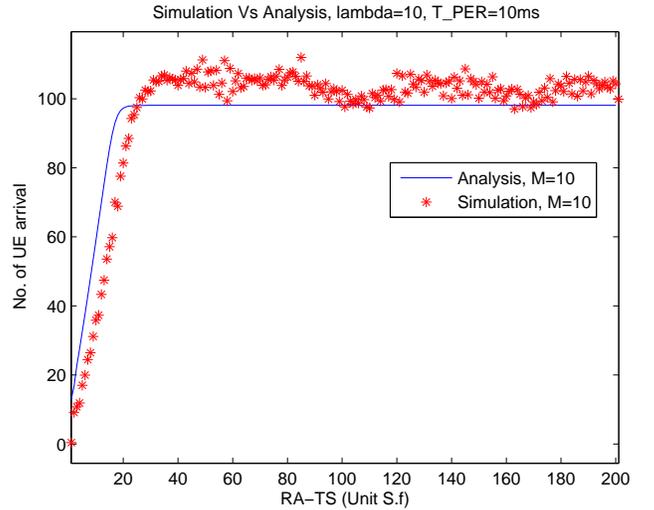}
\caption{Estimated number of H2H UEs arrival after retransmission ($Z_{i}^{H2H(n)}$) and their new arrival ($Z_{i}^{H2H(1)}$) 
vs. simualtion ($W=10$).}
\label{Z_H2H}
\end{figure}
Figures (\ref{Z_m2m}), (\ref{Z_H2H}) demonstrate the number of M2M and H2H arrival vs. their new arrival over RA-TSs. Where we compared the analysis results (i.e., K-Mc model Eq.(\ref{Z})) and simulation. As it comes from these figures, the actual number of active M2M/H2H users is very well matched with the analytical model. From Fig. (\ref{Z_m2m}) one can notice that although M2M new arrivals is modeled with a Beta distribution, however, since the back-off procedure is a uniform distribution over RACH slots, the number of M2M arrivals after $W$ times retransmission is a linear product of $W*Z_{i}^{M2M(1)}$ in steady state. This phenomena is also applicable on H2H users with Poisson arrival, where $\lambda=10$ is the average rate of H2H new arrivals, which reaches to $\lambda*W=100$ per RA-TS in steady state. \\
In Fig. (\ref{PR_success_DA}) and (\ref{PR_success_JA}) we represent the results of access success probability for M2M users, ($P_{s}^{M2M}$), with respect to DA and JA method respectively, where the average rate of H2H calls per second is $\lambda=0.5$. We have verified the same trend for $P_{s}^{M2M}$ for range $0.1\leq\lambda\leq 1$. As it comes from the Fig. (\ref{PR_success_DA}), the probability of access success degrades with respect to the number of preamble allocation to H2H users, $x$. As $x$ increases, $M-x$ lesser number of preambles will be allocated to the M2M users. Where in JA method (Fig. (\ref{PR_success_JA})), the success access probability falls down with a better slope. The reason behind this result is that in JA algorithm the number of $x$ preambles is being shared with H2H and M2M users to content. As the result the M2M users can get success in lesser number of preamble transmission with JA algorithm than in DA (please see Fig. (\ref{preamble_cdf})). 
\begin{figure}[t]
\centering
\includegraphics[scale=0.65]{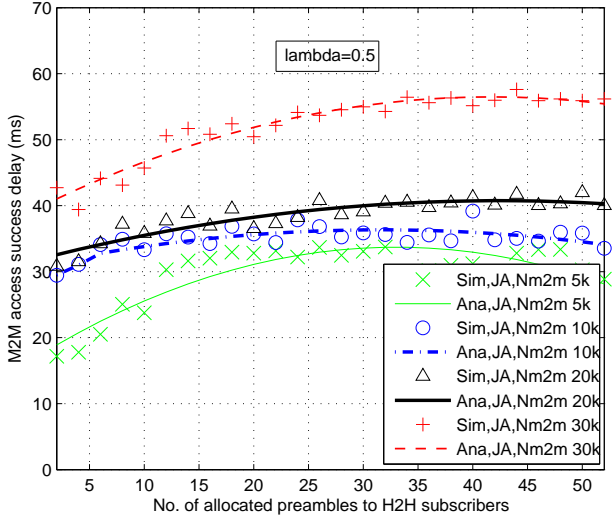}
\caption{Access success delay for M2M calls using JA method, ($\tau^{M2M}_{JA}$).}
\label{delay_JA}
\end{figure}
Following the results from the access success probabilities, we can expect a better access success delay in JA method in comparison to the DA method. Which, as it is depicted in Fig. (\ref{delay_JA}) the results of access success delay with respect to JA method is a smooth convex function with regards to $x$. Where, in Fig. (\ref{delay_DA}) the success delay for M2M users is dramatically increasing by allocating more number of preambles to the H2H users. For computing the success access delay for users, one should note that this value is composed of the delay of departure and delay in receiving success RAR acknowledgement for Msg2 from eNB in RA procedure. In this paper, we have assumed that the delay of other handshaking messages (Msg.3-4) in RACH for allocating the physical uplink channels is constant for all type of users. Therefore, the required success access delay threshold, $\phi$, for M2M users is considered for successfully receiving the RAR message after sending the preamble request (Msg.2). 

\begin{figure}[t]
\centering
\includegraphics[scale=0.6]{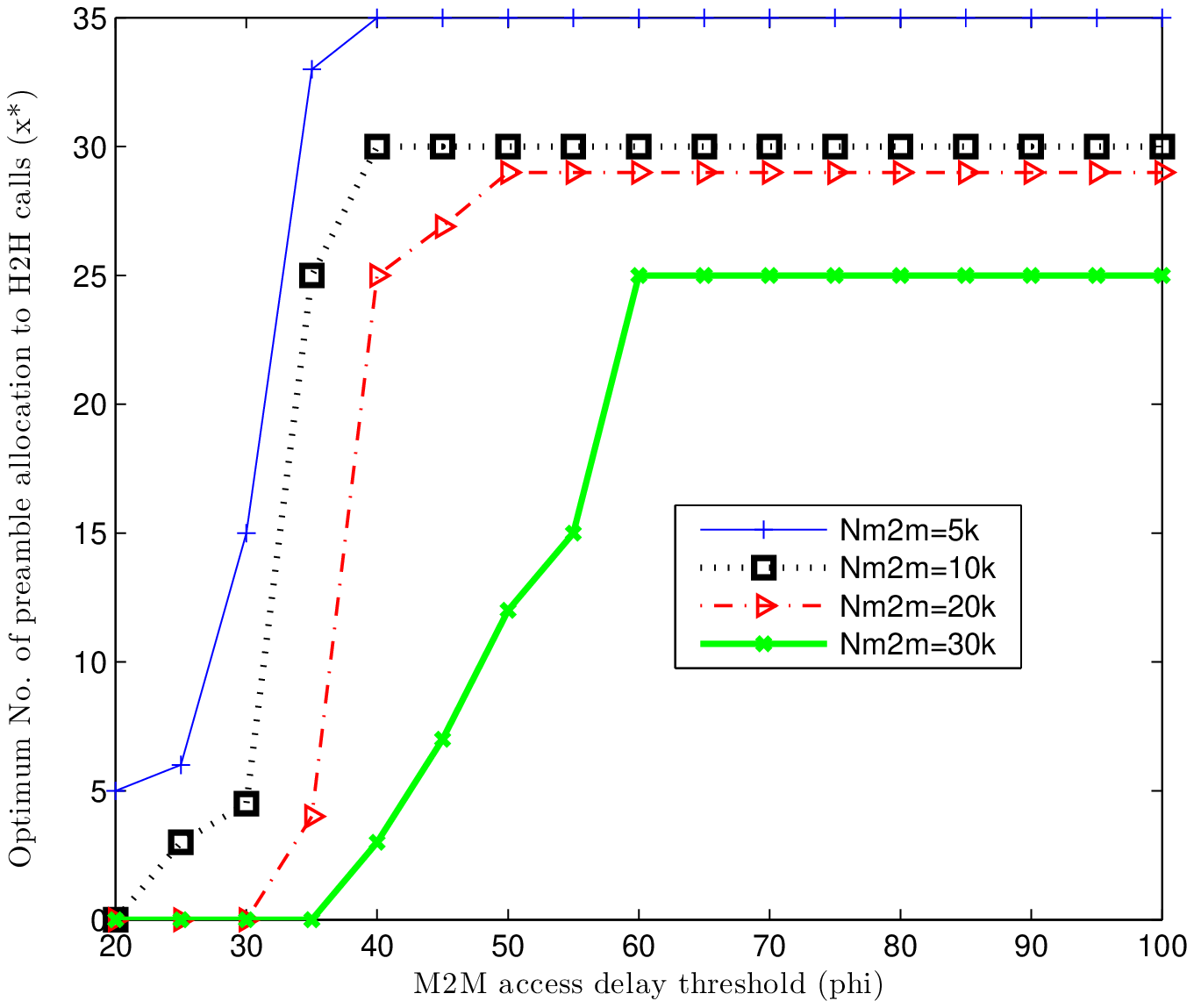}
\caption{Optimum no. of preamble allocation to H2H users, ($x^{\dagger}(\phi)$)}, with respect to the given M2M access success threshold $\phi$.
\label{optimumX}
\end{figure}

The results of optimum preamble allocation to H2H users, $x^{\dagger}$, with  respect to $\phi$ is shown in Fig. (\ref{optimumX}). As we can deduce from this figure, the optimum number of allocating preambles to H2H users per RA-TS decreases with lesser access delay threshold requirement, $\phi$. For example, if there are $N_{MTC}=5000$ number of machine devices in a cell have delay requirement of $\phi\leq20ms$, only $x=5$ number of preambles out of $M$ available preambles in a RA-TS should be allocated to H2H and be shared with M2M subscribers.  
If we consider a real scenario in which there are $N_{MTC}=30000$ machine type devices with 100ms success access delay, which requires receiving the RAR Msg.  in less than 40ms\footnote{Considering the delay of receiving Msg.2 consumes 1/3 of total access success delay in RA procedure.}, there will be only $x=3$ preambles to allocate to the H2H users. Under the bulky arrival of M2M users $N_{MTC}=30000$ with average arrival rate of H2H users $0.1\leq\lambda\leq1$, M2M users can get success access to the eNB only after 35ms. 
\section{Conclusion}\label{section VII}
In this work, we have proposed a closed-form Markovian model for splitting preamble allocation RAN-overload algorithms JA and DA methods. Under the same model we have derived a closed-form analytical expression for computing performance metrics in random access procedure including collision probability, access success probability, access success delay, CDF of preamble transmission and CDF of access success delay for H2H and M2M users. In addition, we examined the JA method for reducing the access success delay for M2M users in the random access procedure. Under the JA algorithm we have suggested a formula for the optimal number of preamble allocation to the H2H subscribers in PRACH. Finally, we have shown that the numerical results from the proposed $K^{th}$ order Markovian model verifies the simulation results.

\end{document}

retransmission